\documentclass[twocolumn]{aastex631}
\usepackage{booktabs}
\usepackage[table]{xcolor}
\usepackage{array}
\usepackage{graphicx}	
\usepackage{amsmath}	
\usepackage{amssymb}    
\usepackage{booktabs}
\usepackage{mathtools}
\usepackage{lipsum} 
\usepackage{xspace}
\usepackage{nolbreaks}

\newcommand{\tauhi}{\tau_{\rm HI}}
\newcommand{\taue}{\tau_{e}}
\newcommand{\Ts}{T_{s}}
\newcommand{\Tcmb}{T_{\rm CMB}}
\newcommand{\Tcmbo}{T_{\rm CMB,0}}
\newcommand{\Tb}{T_{b}}
\newcommand{\xhi}{x_{\rm HI}}

\newcommand{\tcm}{21\,cm }  

\shorttitle{Interpreting Global \tcm Signals with the Radio CMB}
\shortauthors{H. Fronenberg}

\begin{document}

\title{Toward a Complete Interpretation of Global \tcm Absorption Signals Using the Radio CMB}

\correspondingauthor{Hannah Fronenberg}
\email{hfronenberg@uchicago.edu}

\author[0009-0002-7172-5546]{Hannah Fronenberg}
\affiliation{Kavli Institute for Cosmological Physics, University of Chicago, 5640 South Ellis Avenue, Chicago, IL, 60637, USA}

\begin{abstract}

The global \tcm signal is among the most direct probes of cosmic dawn, encoding information about when the first stars formed and how efficiently they heated their surroundings. However, such measurements are limited by bright Galactic foregrounds, ionospheric effects, and instrumental systematics. Even a confident detection would face a further obstacle: the brightness temperature depends on the spin temperature of the gas and the temperature of the radio background only through a single combination, so anomalous signals such as the EDGES \citep{Bowman2018} absorption trough admit both cold gas and excess radio background interpretations. Recently, \citet{Holder2026} showed that resonant \tcm absorption by high-redshift neutral hydrogen imprints an excess optical depth on CMB anisotropies at radio wavelengths, which depends on the spin temperature alone. Here we propose combining a measurement of the CMB hydrogen optical depth with the global signal to break this degeneracy, isolating the radio background temperature without the need for astrophysical modelling. Forecasting four interferometric configurations, we find that a compact $55 \times 55$ array of 6~\,m dishes observing for 4000 hours can establish an EDGES-like optical depth at $4.2\sigma$, while fiducial cosmic dawn histories require roughly an order of magnitude more collecting area. This constitutes a model-independent route to interpreting any global \tcm detection.

\end{abstract}


\section{Introduction}

Cosmic dawn, the era during which the first stars formed and began to reshape their surrounding medium, remains almost entirely unobserved. Prior to the formation of the first luminous sources, the gas cools adiabatically and its kinetic temperature falls below the temperature of the radio background, $T_R$. Lyman-$\alpha$ photons from those first luminous sources then couple the spin temperature of neutral hydrogen, $T_s$, to the gas temperature. The global \tcm signal, $T_b$, is a measurement of the sky-averaged temperature contrast between $T_s$ and $T_R$ weighted by the hydrogen optical depth. As a result, the cosmic dawn global signal is predicted to be an absorption feature, the shape, depth, and width of which encodes information about when the first stars formed, the strength of the Lyman-$\alpha$ coupling, and how efficiently these first astrophysical sources subsequently heated the surrounding medium \citep{Pritchard_Loeb_2012, Liu_Shaw_2020}. 

To date, the Experiment to Detect the Global EoR Signature (EDGES) has reported the only detection: a flattened absorption profile centred at 78\,MHz with a full width at half maximum of $19$\,MHz and an amplitude of $0.5^{+0.5}_{-0.2}$\,K, deeper than the strongest standard predictions by over a factor of two \citep{Bowman2018}. The Shaped Antenna measurement of the background RAdio Spectrum 3 (SARAS 3) experiment rejects the best-fit EDGES profile at 95.3\% while other experiments such as Probing Radio Intensity at high-Z from Marion (PRIZM; \citealt{PRIZM_2019}), the Radio Experiment for the Analysis of Cosmic Hydrogen (REACH; \citealt{REACH_2022}) and Mapper of the IGM Spin Temperature (MIST; \citealt{MIST_2024}) are also undertaking this measurement.

Despite multiple experimental efforts, the sparse detection landscape is no coincidence; global signal measurements are notoriously challenging. At low radio frequencies, ionospheric effects can impart time-variable spectral structure \citep{Datta_2016, Shen_2021}, Galactic synchrotron emission exceeds the signal by four orders of magnitude and must be separated from the data \citep{Bernardi_2015}, and chromaticity in the antenna beam and the ground plane can introduce features that mimic an absorption trough \citep{Mozdzen_2016, Hills_2018, Bradley_2019, Sims_2020}. But suppose these observational challenges are resolved, a challenge that still remains is: how does one fully interpret a global signal measurement? 

The difficulty is that $T_b$ depends on both the spin temperature as well as the radio background temperature, so inferring either one separately carries a significant modelling burden. For example, the anomalously deep trough reported by EDGES spawned two classes of models: one that generates the trough by cooling the gas below the standard adiabatic limit \citep{Barkana_2018}, and another that invokes additional sources to raise the radio background above the CMB temperature \citep{Feng_Holder_2018, EwallWice_2018}. In order to break this degeneracy, other datasets are required, and indeed, the latest upper limits from measurements of \tcm fluctuations by the Hydrogen Epoch of Reionization Array (HERA; \citealt{HERA2026}) have begun to weigh in. Combined with the earlier limits from \cite{HERA2023}, models invoking an excess radio background tend to disfavour the EDGES trough. The authors caution, however, that this inference is strongly dependent on the model priors. Nearly all of the constraining power comes from lower redshift bins that is then extrapolated to cosmic dawn only through each model's assumed smooth thermal evolution. The most model-agnostic limits remain consistent with both excess cooling and an excess radio background, and would require an order of magnitude improvement at $z\sim17.5$ to begin excluding EDGES. Nonetheless, this demonstrates the value of bringing together independent probes, and motivates the search for one that constrains $T_s$ directly, without depending on a model of the underlying astrophysics.

Fortunately, \citet{Holder2026} showed that resonant \tcm absorption by high-$z$ neutral hydrogen imposes an additional optical depth, $\tauhi$, to the CMB at radio wavelengths, in excess of the frequency-independent Thomson optical depth, $\taue$. This results in excess attenuation of CMB anisotropies by an additional factor of $e^{-\tauhi}$ only at radio frequencies. As shown in \mbox{\cite{Holder2026}}, $\tauhi \propto \xhi/\Ts$, where $\xhi$ is the neutral fraction, therefore a measurement of this excess suppression to the CMB fluctuations before reionization, when $\xhi \sim 1$, directly traces the hydrogen spin temperature. In this paper, we propose using measurements of $\tauhi$ in combination with global signal measurements to break the $Ts$--$T_R$ degeneracy and move toward a full, model-independent interpretation of the global signal.

The remainder is organized as follows. In Section~\ref{sec:formalism} we lay out the formalism relating $\tauhi$ to the global signal stating all required assumptions. These two observables are examined across a range of cosmic dawn scenarios, from fiducial histories to the deep trough claimed by EDGES. Section~\ref{sec:forecasts} describes our forecast setup: the array configurations considered, the noise model, and the detectability criteria. In Section~\ref{sec:results} we present forecasts for a range of observational setups and cosmic dawn scenarios, and discuss the observational challenges these measurements face, including foreground contamination and possible mitigation strategies. Section~\ref{sec:discussion} provides an in depth discussion of the constraints this measurement would place on both cold gas and excess radio background models. We conclude in Section~\ref{sec:conclusion} with an assessment of near-term feasibility. Throughout we assume a Planck 2018 cosmology \citep{Planck_2018_IV}.

\section{Formalism} \label{sec:formalism}

The \tcm line is the result of the hyperfine transition of neutral atomic hydrogen. The spin temperature, $T_s$, describes the relative occupancy of the excited spin state to the ground state of this hyperfine transition. Throughout cosmic history, various physical processes can couple the spin temperature to either the kinetic gas temperature, $T_K$, or the CMB temperature, $T_{\rm{CMB}}$. Studying the complex redshift evolution of the spin temperature is a rich physical quantity that sheds light on early expansion history, and on early stellar and galaxy formation (for a detailed review see Section 2 of \citealt{Liu_Shaw_2020}). 

The spin temperature is related to the optical depth of neutral hydrogen given by 

\begin{equation}
  \tauhi \simeq 0.00847\,C_{\rm{cosmo}}(1+z)^{3/2}\,\frac{\xhi}{\Ts}
  \label{eq:tau}
\end{equation}
where $x_{\rm{HI}}$ is the neutral fraction of hydrogen \citep{Furlanetto2006}. The cosmological factor is given by  
\begin{equation}
     C_{\rm{cosmo}} =  \left(\frac{\Omega_b h^2}{0.02237}\right)
  \left(\frac{1-Y_p}{0.7546}\right)
  \sqrt{\frac{0.1424}{\Omega_c h^2 + \Omega_b h^2}},
\end{equation}
where $\Omega_b$ is the baryon energy density, $\Omega_c$ is the cold dark matter energy density, and $Y_p$ is the primordial helium mass fraction. This quantity is scaled such that $C_{\rm{cosmo}} = 1$ in the \textit{Planck} 2018 cosmology which we assume throughout \citep{Planck2018VI}. 

Both the spin-temperature and the hydrogen optical depth are related to the \tcm global differential brightness temperature (often referred to as the global signal), 

\begin{equation}
  \Tb(z) = \frac{\tauhi}{1+z}\,\big(\Ts - T_R\big), 
  \label{eq:tb}
\end{equation}
which is a measure of the temperature contrast between the spin temperature, $T_s$, and the radio background temperature, $T_R$. 

Assuming \textit{no excess radio background} (i.e. $T_R = \Tcmb$), a global signal measurement along with a measurement of the present day CMB temperature, $\Tcmbo$ (e.g. from the Far Infrared Absolute Spectrophotometer; \citealt{Fixen2009}) provides a prediction for the hydrogen optical depth,

\begin{equation}
  \hat{\tau}_{\rm HI} = \frac{0.00847C_{\rm cosmo}\,(1+z)^{1/2}\,\xhi - \Tb}{\Tcmbo}. 
  \label{eq:invert}
\end{equation}

\begin{figure}[t]
    \centering
    \includegraphics[width=\linewidth]{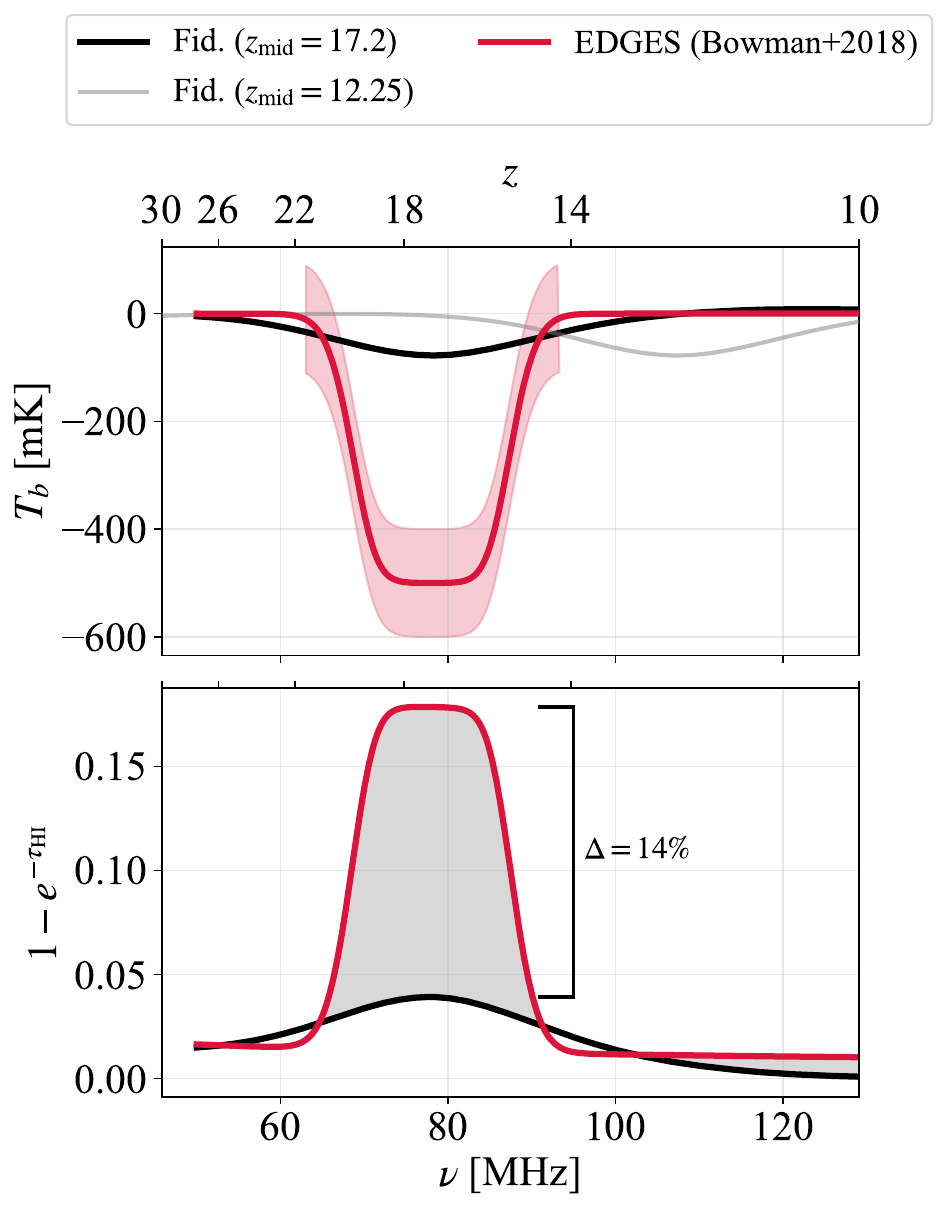}
    \caption{The global \tcm signal and the CMB attenuation it implies. Top: brightness temperature $\Tb$ for two fiducial \textsc{21cmfast} models with absorption troughs of standard depth centred at $z = 17.2$ (black) and $z = 12.25$ (grey), compared with the EDGES best-fit profile (red), which is deeper than either by more than a factor of two. Bottom: the corresponding fractional suppression of CMB anisotropies, $1 - e^{-\tauhi}$, obtained from each spin temperature history via Eq.~\ref{eq:tau}. A measurement of this suppression  determines $\Ts$ independently of $T_R$, breaking the degeneracy inherent in $\Tb$ alone.}
    \label{fig:Brighness_temp}
\end{figure}

Pre-reionization (i.e. $z \gtrsim 12$), when the Universe is effectively neutral, a robust prediction for $\tauhi$ can be made. This optical depth, as shown in \cite{Holder2026}, adds an additional suppression of $e^{-\tauhi}$ to the CMB anisotropies on all scales, analogous to the usual electron optical depth due to the Thomson scattering of CMB photons off of free electrons along the line of sight. The top panel of Figure~\ref{fig:Brighness_temp} shows the brightness temperature for two fiducial \textsc{21cmfast} models \citep{Mesinger2011, Murray2020}, with absorption troughs of comparable, standard depth centred at $z = 17.2$ (black) and $z = 12.25$ (grey), the former chosen to coincide with the EDGES band. Shown in red for comparison is the EDGES best-fit profile \citep{Bowman2018}, which is deeper than either by more than a factor of two. For the CMB temperature, we adopt the current best fit value of 2.725 K and do not include its negligible error contribution. The bottom panel shows the corresponding fractional suppression of CMB fluctuations, $1 - e^{-\tauhi}$. Relative to the fiducial histories, the EDGES profile implies an additional suppression of $14\%$ at $z = 17.2$ and $17.5\%$ when cosmic dawn occurs slightly later (grey).

These differences are informative because, at cosmic dawn, $\tauhi$ responds only to the spin temperature and carries no dependence on the radio background. A measurement of $\tauhi$ consistent with the prediction for the EDGES profile given by Eq.~\ref{eq:tau} would not only corroborate the detection, but also confirm that the gas is anomalously cold, disfavouring interpretations that invoke an excess radio background. Indeed, recent work has shown that cold gas models are favoured over those invoking an excess radio background \citep{Cang2025}. Conversely, if a deep $\Tb$ absorption trough were observed and the corresponding prediction for $\tauhi$ is \textit{not} observed, this would instead point to an excess radio background as the dominant explanation for the EDGES trough. In any case, $\tauhi$ serves as a highly complementary and degeneracy breaking counterpart to the global signal. In the following section, we lay out the framework needed to assess whether the differences between models in Fig.~\ref{fig:Brighness_temp} are measurable. We define the observables, present various instrument and noise models, and establish criteria for detection.

\section{Forecast Scenarios}
\label{sec:forecasts}

To evaluate the feasibility of detecting $\tauhi$, we adopt the observational strategy proposed in \cite{Holder2026}. Because the CMB anisotropies are exquisitely well measured at millimetre wavelengths, existing maps provide a template with negligible uncertainty over the multipole range of interest, and the measurement reduces to constraining the amplitude of that template in a radio map. The spatial structure in the radio maps are expected to be the same as that in the millimetre maps, suppressed by $\alpha = e^{-\tauhi}$. The radio CMB maps can therefore be modelled by 

\begin{equation}
    T^{\rm radio}(\mathbf{\hat{n}, \nu}) = \alpha(\nu)  T^{\rm mm}(\mathbf{\hat{n}}) + n(\mathbf{\hat{n}, \nu})
\end{equation}
where $T^{\rm radio}$ is the observed radio map, $T^{\rm mm}$ is the CMB template map in the millimeter, and $n(\mathbf{\hat{n}, \nu})$ is the noise. The minimum variance estimator $\hat{\alpha}$ for the template amplitude is 

\begin{equation}
    \hat{\alpha}(\nu) = \frac{\sum_{u,v} \tilde{T}^{\rm mm,\dagger}(u,v)\sigma^{-2}(u,v)\tilde{T}^{\rm radio}(u,v)}{\sigma^{-2}(u,v)|\tilde{T}^{\rm mm}(u,v)|^2}
\end{equation}
where tilde denotes the Fourier transform and $u$ and $v$ are the Fourier dual to the sky coordinates in the flat sky approximation. Assuming the radio map is made by a radio interferometer, the error associated with this estimator is 

\begin{equation}\label{eq:delta_alpha}
    \Delta\hat{\alpha}(\nu)= \frac{T_{\rm sys}}{\sqrt{\Delta\nu \Delta t}} \left(\sum _{uv} C_{\ell=2\pi|\mathbf{u}|}N_{uv} \right)^{-1/2} 
\end{equation}
where $T_{\rm sys}$ is the system temperature, $\Delta\nu$ is the bandwidth, $\Delta t$ is the total observing time, and $N_{uv}$ is the $uv$-coverage (i.e. number of times each $uv$-mode is measured by the interferometer)\footnote{Also, see \cite{Holder2026} for a simplified expression of the error, $\Delta\hat{\alpha}$, where a tightly packed square array is assumed. Given that these forecasts are done at low frequencies where the approximation is expected to break down, we do not employ this approximation here.}. The CMB angular power spectrum, $C_\ell$ is obtained from the publicly available code \textsc{camb}\footnote{\url{https://github.com/cmbant/CAMB}}.

We consider four observational scenarios that all assume $f_{\rm{sky}} = 0.1$, $t_{\rm{obs}} = 4000$ hours, and $T_{\rm{sys}} = 180 (\nu/180\,\rm{MHz})^{-2.55}$ K (i.e. sky dominated system temperature; \citealt{pygsm2008, pygsm2016,Furlanetto2006}). For $60 \lesssim \nu \lesssim 120$, this results in an exceptionally bright system temperature between $\sim 1000-2000$ K. These scenarios vary in their dish size and element configuration:

\begin{itemize}
    \item \textbf{HERA-like}: Since HERA has a 320 element core, we approximate a HERA-like instrument as an $18 \times 18 $ array of 14 m dishes. 
    
    \item \textbf{Threshold Array}: This configuration is sized for a marginal ($\sim1\sigma)$ discrimination between EDGES and the fiducial cosmic dawn model. This is a $20\times20$ array of 6 m dishes. 
    
    \item \textbf{Dedicated (Holder+2026)}: This array corresponds to the proposed dedicated array from \cite{Holder2026} which is a $30\times30$ array of 6 m dishes.
    
    \item \textbf{Robust Array}: This $55 \times 55$ array of 6 m dishes, in contrast to the Threshold Array scenario, is the array required to robustly discriminate ($>3\sigma$) between the EDGES and other models, or to make marginal detections of the fiducial models. 

\end{itemize}

We adopt 6\,m dishes for the three purpose-built configurations, matching the fiducial array of \citet{Holder2026} so that our forecasts are directly comparable. This choice is also favoured by their Eq.~11, which weakly prefers small elements: the shortest available baseline is set by the dish diameter, and the CMB anisotropies we seek to measure are strongest on the largest angular scales. At these frequencies, however, smaller dishes are only a few wavelengths across, and we caution that the idealized treatment adopted here, in which elements are taken to be uncorrelated and their beams described by ray optics, is unlikely to hold in detail. A realistic forecast would require full electromagnetic simulation of the array, but as \citet{Rath2025} point out, numerical artifacts \citep{Mahesh2021} and large computational costs render this approach impractical for this initial sensitivity estimate. Mutual coupling between closely packed dishes is a known and serious systematic for compact arrays of this kind \citep{Kern2019,Kern2020, Josaitis2022, Rath2025}, and has recently become the dominant systematic for HERA \citep{HERA2026}. We therefore regard our sensitivity estimates as optimistic, and return to the implications in Section~\ref{sec:results}.

In practice our 6\,m dish configurations do not probe $\ell < 10$ (with $\ell \approx 2\pi|\mathbf{u}|$), the compact arrays considered here have $\ell_{\min} \gtrsim 10$ across the band. This is just as well. Equation~\ref{eq:delta_alpha} might suggest that the lowest multipoles are the most valuable, since the CMB anisotropies are brightest on the largest angular scales, but those expressions assume a template of negligible uncertainty. At the very lowest multipoles, the \textit{Planck} maps cannot be treated this way: residual Galactic foregrounds and large-scale systematics become significant below $\ell \sim 10$, so the template carries non-negligible error of its own. We therefore caution against extrapolating to larger angular scales than those sampled here.

With these details in mind, the $uv$-coverage, $N_{uv}$, for each configuration is generated using the publicly available code \textsc{LIMstat} \citep{Fronenberg_Liu_2024}\footnote{\url{https://github.com/McGill-Cosmic-Dawn-Group/LIMstat/tree/main}}. In each case we assume a noise-limited measurement and defer the question of foreground contamination and removal for a future end-to-end analysis, though we comment on the role of foregrounds in the sections that follow. 

In the results section below, we quote two figures of merit. The first is the the detection significance with which the template amplitude is measured at all, 

\begin{equation}\label{eq:sn-det} 
  \sigma^2_{\rm det}
  = \sum_\nu \left(\frac{\hat{\alpha}(\nu)}{\Delta\hat\alpha(\nu)}\right)^2
\end{equation}
and the second is the significance with which a given model can be separated from a null hypothesis,

\begin{equation}
\sigma^2_{\rm null}
  = \sum_\nu \left(\frac{\hat{\alpha}_{\rm null(\nu)} - \hat{\alpha}_{\rm model}(\nu)}{\Delta\hat\alpha(\nu)}\right)^2.
  \label{eq:sn-dist}
\end{equation}
Here $\alpha_{\rm null}$ may be the prediction of a competing history, or, as we adopt throughout, $\alpha_{\rm null} = 1$, corresponding to an unattenuated CMB with no excess optical depth from cosmic dawn. 


\section{Results}
\label{sec:results}

Table~\ref{tab:detections} summarizes the figures of merit for each observing scenario. In every case the cumulative significance is calculated over 19\,MHz, approximately the FWHM of the EDGES trough.

\begin{figure}[t]
    \centering
    \includegraphics[width=\linewidth]{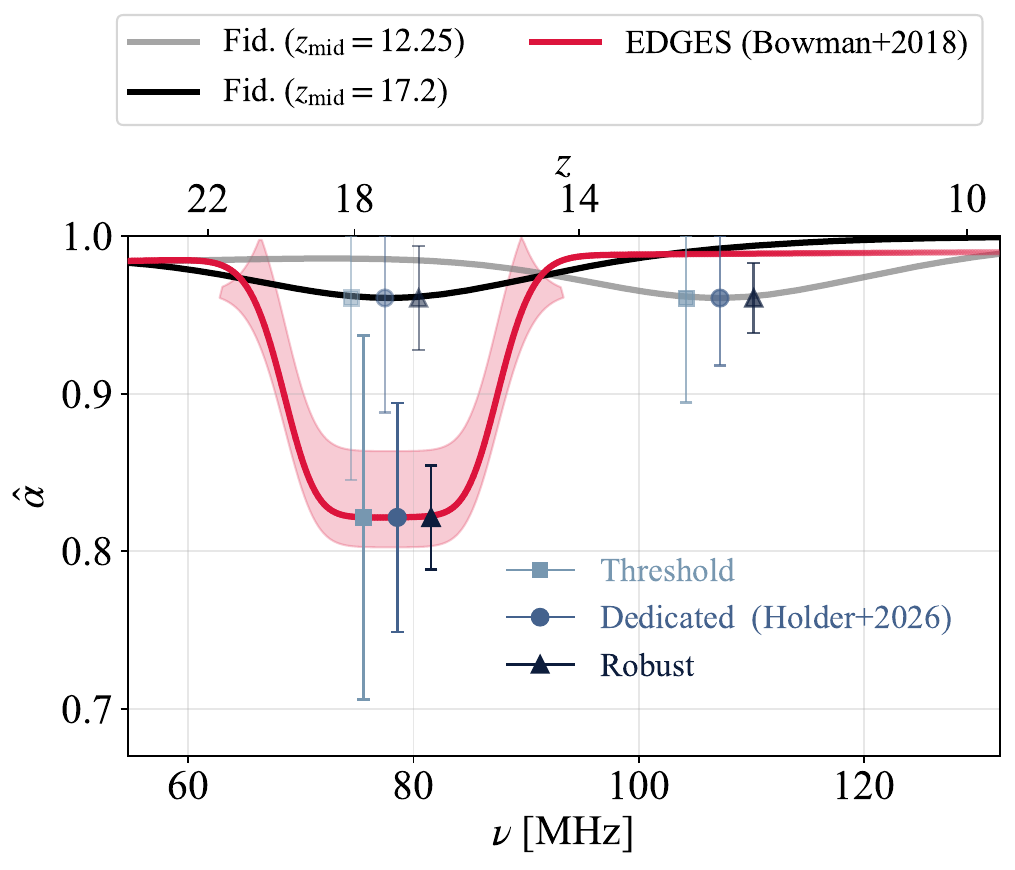}
    \caption{Forecast constraints on the template amplitude $\hat{\alpha}$ for three observational setups: Threshold (squares), Dedicated (circles; \citealt{Holder2026}), and Robust (triangles). Constraints are obtained for three spin temperature histories chosen to bracket the plausible parameter space: the EDGES profile (deepest trough), and fiducial histories with standard-depth troughs at $z = 17.2$ (early cosmic dawn) and $z = 12.25$ (late). The Robust configuration confirms an EDGES-like history at better than $4\sigma$, while the fiducial histories reach $\sim 2\sigma$ at $z = 12.25$ and $\sim 1\sigma$ at $z = 17.2$. The anomalous scenarios are thus the most readily identified. The deeper the trough, the larger $\tauhi$ and the more sharply the models separate.}
    \label{fig:Forecast_sensitivity}
\end{figure}

A HERA-like array detects the radio CMB at $3.03\sigma$ for an EDGES-like history and up to $6.00\sigma$ for the lower redshift fiducial model, but cannot constrain any of the models individually. The limitation is the dish size rather than the element count. The shortest baseline is set by the $14\,$m dish diameter, and therefore the array does not sample the large angular scales that carry the most power. The Threshold configuration achieves a higher detection significance by over a factor of two with only 23\% more elements. This sensitivity gain therefore comes almost entirely from the reduction in the dish diameter from 14\,m to 6\,m, enabling the detection of large scale CMB modes. With this increased sensitivity, the Threshold array, sized for marginal discrimination of the EDGES history, yields $\sigma_{\rm null} = 1.21$ while detecting the anisotropies at $7.11\sigma$. It is, however, unable to place constraints on the fiducial scenarios as seen in Figure~\ref{fig:Forecast_sensitivity}. 

The Dedicated array of \citet{Holder2026} is able to detect the EDGES model at almost $2\sigma$, while the Robust configuration reaches $4.23\sigma$, securing an EDGES-like $\tau_{\rm HI}$ while also placing marginal constraints on the fiducial histories at $1.19\sigma$ at $z=17.2$ and $1.77\sigma$ at $z = 12.2$. The two fiducial models share an identical $\tau_{\rm HI} = 0.04$, yet the later one is constrained substantially better in every configuration. Mapping to observing frequency, $z = 17.2$ and $z = 12.25$ correspond to 78.0 and 107.2\,MHz respectively, and the improvement follows almost entirely from the decreased sky temperature at higher frequencies.

\begin{deluxetable}{lcccccc}
\tablecaption{Forecast significances for each observing scenario and
  $\tau_{\rm HI}$ model. \label{tab:detections}}
\tablehead{
  \colhead{} &
  \multicolumn{2}{c}{EDGES} &
  \multicolumn{2}{c}{Fid. ($z=17.2$)} &
  \multicolumn{2}{c}{Fid. ($z=12.25$)} \\
  \colhead{} &
  \multicolumn{2}{c}{($\tau_{\rm HI}=0.197$)} &
  \multicolumn{2}{c}{($\tau_{\rm HI}=0.04$)} &
  \multicolumn{2}{c}{($\tau_{\rm HI}=0.04$)} \\
  \cline{2-3} \cline{4-5} \cline{6-7}
  \colhead{} &
  \colhead{$\sigma_{\rm det}$} & \colhead{$\sigma_{\rm null}$} &
  \colhead{$\sigma_{\rm det}$} & \colhead{$\sigma_{\rm null}$} &
  \colhead{$\sigma_{\rm det}$} & \colhead{$\sigma_{\rm null}$}
}
\startdata
HERA      & 3.03 & 0.51 & 3.54 & 0.14 & 6.00 & 0.25 \\
Threshold & 7.11 & 1.21  & 8.32 & 0.34 & 14.51 & 0.59 \\
Dedicated & 11.29 & 1.92 & 13.21 & 0.54 & 22.49 & 0.97 \\
Robust    & \textbf{24.94} &\textbf{ 4.23} & \textbf{29.18} & \textbf{1.19} & \textbf{43.28} & \textbf{1.77} \\
\enddata
\tablecomments{ $\sigma_{\rm det} \equiv \alpha/\Delta\alpha$ is the significance of detecting the CMB at radio frequencies (null hypothesis: $\alpha=0$); $\sigma_{\rm null} \equiv (1-\alpha)/\Delta\alpha$ is the significance of
 nonzero excess optical depth (null hypothesis: $\alpha=1$).}
\end{deluxetable}

\begin{figure}[t]
    \centering
    \includegraphics[width=\linewidth]{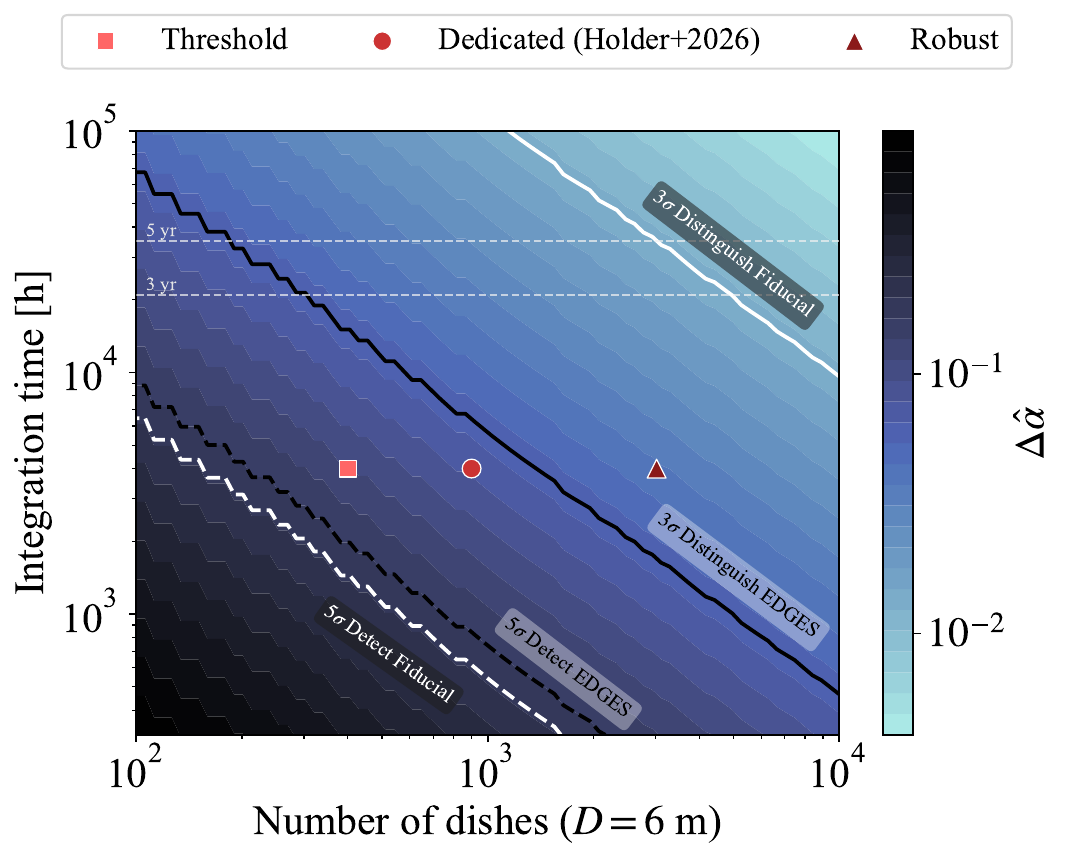}
    \caption{Required sensitivity $\Delta\hat{\alpha}$ as a function of integration time and number of 6\,m dishes. Dashed contours mark the $5\sigma$ threshold for detecting the radio CMB at all, that is, rejecting the null hypothesis of no signal consistent with the CMB template. Solid contours mark the more demanding $3\sigma$ threshold for distinguishing a given spin temperature history from the alternatives. Both are shown for the EDGES profile (black) and the fiducial history (white). Detection comes slightly sooner for the fiducial case, since weaker suppression leaves $\alpha$ closer to unity, but the ordering reverses for discrimination: the EDGES threshold is reached far earlier because the model is so deep and extreme, while constraining the subtler fiducial history demands substantially more collecting area and integration time. The radio CMB is therefore an accessible near-term target in its own right, with $\tauhi$ science following as the array grows.}    
    \label{fig:Instrument_builder}
\end{figure}

Figure~\ref{fig:Instrument_builder} plots the sensitivity in terms of $\Delta\hat{\alpha}$ as a function the number of 6\,m dishes and the integration time, evaluated at 78\,MHz. Dashed contours mark the $5\sigma$ threshold for detecting the radio CMB, solid contours mark the $3\sigma$ threshold for establishing excess optical depth, shown for the fiducial history (white) and the EDGES history (black). Horizontal lines are overlaid corresponding to three and five observing years, taking an observing year to be 80\% of a calendar year.

The ordering of the two sets of contours inverts between the two tests. An initial detection is marginally harder for the EDGES thermal history, since the deeper trough suppresses the amplitude of CMB anisotropies more than other conservative models. However, the trend reverses when it comes to the experimental setup required to actually constrain these models. The deeper EDGES trough is easier to robustly constrain since it is much more distinct from the null hypothesis of $\alpha = 1$. Reaching a $3\sigma$ constraint of an EDGES-like $\tau_{\rm HI}$ can be achieved within 5 years with 1000 dishes or fewer, whereas the same significance on the fiducial history would require on the order of $10^4$ dishes. The scenario that is initially harder to detect is subsequently by far the cheapest to identify.

The results in Figure $\ref{fig:Instrument_builder}$ also reveal a scalable experimental strategy. Detection of the radio CMB is within reach of current generation arrays. As construction continues, the array passes through a regime in which an anomalously cold history can be confirmed or disfavoured at high significance before reaching the scale at which fiducial thermal histories can also be constrained. Each stage yields a scientifically meaningful statement, and no stage is wasted effort toward the next.

The scale of the full Robust configuration is comparable to that of existing next-generation facilities. At $3025$, $6\,$m dishes, it requires a similar element count and essentially the same dish diameter as the Deep Synoptic Array (DSA, formerly DSA-2000; \citealt{DSA_2000}), which will comprise $1650 \times 6.15\,$m dishes. Despite their similarity in the number of elements, the two differ in a multiple ways that favour the array proposed here: the close-packed configuration spans only a few hundred metres rather than tens of kilometres required for the DSA, substantially reducing infrastructure cost, and observing at an order of magnitude lower frequency places the system in the sky-noise-dominated regime while reducing the instantaneous bandwidth, and hence the correlator and data-transport requirements, by a significant factor.

Despite the promise of these measurements, significant challenges remain, foremost among them is foreground contamination. At $\sim$100\,MHz the Galactic synchrotron emission exceeds the CMB anisotropies by six to seven orders of magnitude in brightness temperature. Foregrounds that are uncorrelated between the radio map and the cleaned \textit{Planck} map contribute variance but no bias, and this variance can be suppressed by cleaning the radio maps. The greater danger is instead the \emph{correlated} contaminants: Galactic synchrotron at $\sim100$\,MHz and at \textit{Planck} frequencies is the same emission, that is, the same cosmic-ray electron population in the same magnetic field, and is therefore morphologically correlated across frequencies. Moreover, on the large angular scales relevant here, even physically distinct components such as thermal dust and anomalous microwave emission (AME) share the large-scale morphology of the Galaxy and cannot be assumed to be entirely uncorrelated with the radio sky. Any such Galactic residual surviving in the \textit{Planck} map may correlate with the enormous radio-frequency synchrotron signal to produce a multiplicative bias that, unlike noise, does not integrate down. Because this bias scales as the product of the residuals in the two maps, the burden of foreground cleaning is shared between them. Radio foregrounds can be suppressed
using well-established techniques that exploit the spectral smoothness of the foregrounds \citep{Liu_Tegmark_2011, Shaw_2014, Shaw_2015, Mertens_2018, Liu_Shaw_2020, Anstey_2021}, an approach demonstrated at these frequencies with large arrays \citep{Eastwood_2018, Eastwood_2019, Mertens_2020}, though care must be taken to calibrate the cosmological signal loss that aggressive cleaning can induce \citep{Paciga_2013, Cheng_2018, Pascua_2025}.

Encouragingly, the \textit{Planck} component-separated maps have already been demonstrated to have low residual correlation with the Haslam 408\,MHz synchrotron template, at the level of a few percent and consistent with chance correlations \citep{Planck_2018_IV}, suggesting that with a sufficient level of foreground cleaning on the radio side, aided by masking of the Galactic plane and explicit deprojection of low-frequency templates \citep{Remazeilles_2015, Dowell_2017}, the measurement could be brought into the detection regime. Establishing the required level of cleaning, and verifying that it is achievable in the presence of realistic instrumental effects, is a central goal of future end-to-end analyses. 

Even with foreground cleaned maps, it is equally crucial to then validate data quality, and luckily, various features of the CMB can provide decisive checks for power-law residuals. For instance, the amplitude of a true CMB detection is achromatic in thermodynamic temperature, whereas any Galactic bias inherits the synchrotron or dust spectral energy distribution. Keeping the radio map fixed and repeating the analysis against individual Planck frequency maps and against the four independent component-separation solutions \citep{Planck_2018_IV} therefore constitutes a null test. Any frequency dependence between the pairs of maps would not only indicate the presence of residuals, but the observed power law scaling can indicate which foreground components are still leaking into the \textit{Planck} maps.

Alternatively, while from a pure sensitivity standpoint only the largest-scale modes are required for a $\tauhi$ detection, the low-$\ell$ CMB spectrum is nearly featureless and thus difficult to distinguish from residual contaminants. Resolving the first acoustic peak, achievable by extending the compact array outward or by installing a small number of outrigger stations, would provide an unambiguous confirmation of the shape of the CMB angular power spectrum. The present analysis neither includes radio foregrounds nor imperfections in the \textit{Planck} map, it therefore remains unclear exactly how much the error budget can absorb due to the additional variance these contaminants introduce. A robust map-level, end-to-end analysis that incorporates realistic radio foregrounds, the chromatic response of the interferometer, and \textit{Planck} residual systematics is therefore the essential next step in establishing the near-term feasibility of this measurement.

\section{Discussion}\label{sec:discussion}

As previously mentioned, the EDGES signal has been difficult to interpret since a deep absorption trough could be the result of anomalously cold gas (i.e. $T_s$ is smaller than expected), of an excess radio background (i.e. $T_{R}$ is larger than expected), or a combination of both effects which raise the effective brightness contrast. Due to this degeneracy, it is difficult to interpret \textit{any} global signal measurement, even one that is consistent with expected thermal histories. Since a measurement of $\tauhi$, is insensitive to the radio background, a measurement of $\tauhi$ provides a completely independent lever arm, breaking this degeneracy from global signal measurements, and providing constraints on both $T_s$ and $T_{\rm CMB}$ at the time of cosmic dawn, or potentially, any excess radio background contribution.

A large body of literature has commented on possible resolutions to the EDGES anomaly. Once class of models explores excess gas cooling as the dominant mechanism that produces the observed trough. In order to cool the gas temperature and therefore  the spin temperature of hydrogen to which it is indirectly coupled via the Wouthuysen-Field effect \citep{Wouthuysen_1952,Field_1959a, Field_1959b}, baryon-dark matter scattering emerges as a possible mechanism \citep{Barkana_2018,Fialkov_2018,Barkana_2018b}. A burgeoning class of models capable of achieving the required level of cooling propose the existence of milli-charged dark matter whose charge is as much as one million times smaller than that of the electron \citep{Munoz_Loeb_2018,Berlin_2018,Kovetz_2018, Slatyer_2018, Munoz_Dvorkin_2018, Liu_2019}. In these scenarios, Coulomb-like scattering with cross section $\sigma \propto \epsilon^2 v^{-4}$ transfers heat from the baryons to the colder dark matter, with a cooling rate that grows with the millicharge $\epsilon$ and the number density of scatterers ($\propto f_{\rm dm}/m_\chi$), such that the gas temperature, and hence the spin-temperature at cosmic dawn, is set by the combination of dark matter mass, charge, and millicharged fraction. Therefore, in addition to the thermal history, measurements of $\tauhi$ by its direct relationship with $T_s$ can also provide novel constraints on these dark matter models. 

The alternative class of EDGES solutions invokes a radio background in excess of the CMB, which enhances the temperature contrast in Eq. \ref{eq:tb} by raising $T_R$ rather than lowering $T_s$. Two sources have been explored: astrophysically sourced (early accreting black holes \citep{EwallWice_2018}, supernovae of Population III stars \citep{Jana_2019}, steepened faint end UV luminosities of early galaxies \citep{Mirocha_Furlanetto_2019}, and radio-loud galaxies \citep{Reis_2020}), and a homogeneous synchrotron-like excess of potentially exotic origin, \citep{Feng_Holder_2018,Fialkov_Barkana_2019,Fraser_2018,Pospelov_2018} motivated in part by the unexplained low-frequency excess seen by ARCADE-2 \citep{Seiffert_2009,Fixen_2011} and the Long Wavelength Array (LWA; \citealt{Dowell_Taylor_2018}). Since $\tauhi$ provides a direct constraint on $T_s$ at early times when the Universe is neutral, indirect constraints can be placed on the radio background when measurements of $\tauhi$ and $T_b$ are combined. It it worth noting that recent observations already pressure the excess background scenario. The SARAS 3 non-detection rejects the EDGES best-fit profile at 95\% confidence \citep{Singh_2022} which translates into limits disfavouring a homogeneous excess above 6\% of the CMB at $1.42\,$GHz and disfavouring a bright radio-galaxy population at cosmic dawn \citep{Bevins_2022}. In addition, a recent reanalysis of the EDGES data finds that the data disfavour a cosmological signal of non-standard depth, attributing the flattened trough to unmodeled systematics rather than an excess background \citep{Cang2025}\footnote{This analysis jointly fits a Pop III radio-galaxy background, foregrounds, and calibration residuals in a joint likelihood framework, so this result is a model-dependent claim.}. 

Ultimately, $\tauhi$ serves as a background-independent thermometer for the intergalactic medium at cosmic dawn, capable of discriminating between exotic cooling, an excess radio background, or a purely systematic origin of not only the EDGES signal, but of any global signal measurement.

\section{Conclusion}
\label{sec:conclusion}

In this paper, we propose using measurements of the excess optical depth, $\tauhi$, imprinted on CMB anisotropies at radio wavelength \citep{Holder2026}, to break the degeneracy that limits the interpretation of \tcm global signal measurements. Since $\tauhi \propto \xhi/\Ts$ has no dependence on the radio background and constrains the spin temperature directly, combining it with measurements of $\Tb$ enables constrains on the radio background temperature with no astrophysical modelling. Forecasting four array configurations, we find that current-generation instruments can achieve a detection of the radio CMB, but constraining particular thermal histories is considerably more demanding. Our Robust configuration consisting of a $55 \times 55$ array of 6\,m dishes observing for 4000 hours constrains and EDGES-like history at $4.23\sigma$ and the fiducial histories only marginally. Somewhat counterintuitively, the deeper the absorption trough, the harder the radio CMB is to detect but the easier the underlying history is to identify. 

The global signal and measurements of $\tauhi$ are complementary in an unusually clean way. It is rather straightforward to reach the thermal sensitivity required to measure the global signal, but the measurement is more difficult to interpret and more difficult to extract, with systematics and foreground modelling being the limiting factors. Conversely, the interpretation of a $\tauhi$ measurement at cosmic dawn is far less ambiguous, but reaching the required sensitivity demands on the order of $10^3$ elements observing for several thousand hours. Where the global signal is systematics-limited, $\tau_{\rm HI}$ is sensitivity-limited. 

With these challenges in mind, several questions remain open. The forecasts presented here do not explicitly model the foregrounds , and although we have argued that having a mm-wave CMB template to match with the radio sky reduces the burden of foreground cleaning, explicit verification is required through end-to-end simulation. In addition, the compact array configurations that are favoured in terms of sensitivity are also the most susceptible to systematics like mutual coupling \citep{HERA2026, Rath2025}. It is not yet clear whether deep reflector geometry such as the one designed for the equally compact Canadian Hydrogen Observatory and Radio transient Detector (CHORD; \citealt{CHORD2026}) can mitigate these systematics entirely. And finally, we have assumed throughout that at these early stages of galaxy formation, the intergalactic medium remains effectively entirely neutral ($\xhi \approx 1$). Relaxing this assumption introduces an additional, if modest, source of uncertainty, and constraints on $\Ts$ would inherit some degree of model-dependence.

Disentangling the thermal and ionization history of the first billion years is a great undertaking and will not be accomplished by any single observable. Instead, the global signal, the \tcm power spectrum, and the hydrogen optical depth each constrain different aspects of the same underlying physics, each being limited by different experimental challenges. It is therefore through their combination that a complete picture of cosmic dawn is most likely to be found.

\begin{acknowledgments}
I would like to thank Gil Holder and Adrian Liu for many insightful discussions and for their work that prompted my interest in this measurement. I thank Jessica Zebrowski, Nicholas Huang, Abigail Vieregg, and Bobby Pascua for their expertise in experimental design and feasibility, and Jack Stethem and Michael Hétu for useful discussions. HF is supported by the Kavli Institute for Cosmological Physics at the University of Chicago, and the Kavli Foundation.
\end{acknowledgments}

\bibliography{refs}{}
\bibliographystyle{aasjournal}

\end{document}